\documentclass[a4paper]{jpconf}
\usepackage{graphicx}
\usepackage{amssymb}
\usepackage{amsmath}
\usepackage{feynmp}
\begin{document}
\title{Unusual identities for QCD at tree-level}

\author{N E J  Bjerrum-Bohr$^1$, Poul H Damgaard$^1$, Bo
Feng$^{2,3}$ and  Thomas S{\o}ndergaard$^1$}

\address{$^1$\small Niels Bohr International Academy and Discovery Center,\\
 \small The Niels Bohr Institute,  \small Blegdamsvej 17, DK-2100
Copenhagen, \small Denmark}
\address{$^2$\small Center of Mathematical Science,
\small Zhejiang University, Hangzhou, \small China}
\address{$^3$\small Kavli Institute for Theoretical Physics China,
\small CAS, \small Beijing 100190, China}

\ead{}

\begin{abstract}
We discuss a set of recently discovered quadratic
relations between gauge theory amplitudes.
Such relations give additional structural
simplifications for amplitudes in QCD. Remarkably,
their origin lie in an analogous set of relations
that involve also gravitons. When certain gluon helicities
are flipped we obtain relations that do not involve
gravitons, but which refer only to QCD.
\end{abstract}

\section{Introduction}

At the Large Hadron Collider at CERN
intense beams of protons are already colliding at historically
high energies.
The amounts of data taken from these collisions are enormous and
it will require massive efforts from both experimentalists and
theorists to uncover signals of new physics above `background'
QCD processes. From the theoretical side the provision of precise 
cross sections for scattering processes is therefore crucial. 
Such computations are, however,
tremendously complex if one employs traditional techniques. Luckily,
new computational methods inspired by both string theory and
twistor variables have been rapidly invented in the last decade. These
novel approaches
have succeeded in providing us a series of quite 
efficient toolboxes for computations. Amazingly, we are nevertheless
still learning more about gauge theories and the surprising hidden
simplifications that such theories give rise to in perturbation
theory. Gauge invariance seems to be the essential ingredient: it
is because different parts of calculations can look complicated
or simple depending on the chosen gauge. From this perspective, 
the complexity of perturbative
calculations in QCD and QCD-like theories is partly fake and
only due to inconvenient choices of computation. It is essential 
to uncover as many relations among {\em gauge invariant
observables} as possible.

In this context it came as a complete surprise to everybody
a few years ago that there could be
new non-trivial identities among color-ordered amplitudes in QCD, the 
relations conjectured by Bern, Carrasco and Johansson \cite{BCJ}. 
This set of BCJ-relations, as they are now called, can be proven
most easily using string theory \cite{BDV,Stieberger}, but an inductive
proof based on quantum field theory alone has also recently been established
\cite{Feng,Chen:2011jx}. What is particularly remarkable about BCJ-relations
is that they are sufficiently restrictive to reduce the basis of
color-ordered amplitudes from size $(n-2)!$ to $(n-3)!$ independently
of the helicity configurations. This follows again very naturally
from string theory \cite{BDV}.

Recently, we have found that
there are yet more relations between such color-ordered amplitudes
in Yang-Mills theories (and Yang-Mills theories coupled to matter)
once we fix helicities \cite{BjerrumBohr:2010zb}. 
These new identities take the form of 
a linear sum of {\em products} of two color-ordered amplitudes in
Yang-Mills theory that, surprisingly, vanish. Looked at from field theory
alone this is a most unusual phenomenon, since quadratic relations
among amplitudes have, as far as we know, 
never been seen before, in any context. The 
reason why they nevertheless appear is, amazingly, related to the
perturbative relationship between gravity amplitudes and 
color-ordered Yang-Mills amplitudes, the so-called KLT-relations
\cite{KLT,KLTshort,BjerrumBohr:2010yc,Feng:2010br,BjerrumBohr:2010hn}.

Here we discuss 
these quadratic relations between Yang-Mills amplitudes 
\cite{BjerrumBohr:2010zb} in some greater detail.
Our discussion is structured in the following way. 
In section \ref{gauge_amp}
we review some well-known properties of tree-level gauge-theory amplitudes. 
In section \ref{quad_rel}
we present the quadratic relations along with several explicit examples, and 
in section \ref{proof}
we sketch the proof of these relations. For full details we refer 
to the paper \cite{BjerrumBohr:2010yc}.
Finally, in section \ref{con} we have our conclusions.

\section{Gauge-theory amplitudes\label{gauge_amp}}
To set up some notation, we begin with a brief review of the current state of 
tree-level gauge-theory amplitudes. This
will also serve as a nice comparison between the amplitude relations that 
were known previosuly and the new ones we will discuss here.

It is well known that
when working with tree-level gauge theory amplitudes $\mathcal{A}_n$, 
it is convenient
to introduce the color-ordered \textit{sub}amplitudes $A_n$. 
They are related to the full color-dressed amplitudes by
\begin{align}
\mathcal{A}_n = \sum_{\sigma\in S_{n-1}} 
\mathrm{Tr}[T^{1}T^{\sigma(2)}\cdots T^{\sigma(n)}] 
A_n(1,\sigma(2),\ldots,\sigma(n)).
\end{align}
Here $T^i$ are the generators of the gauge group, and the sum runs over 
all permutations of leg $2,3,\ldots,n$.
In this way one has completely decoupled the color-structure from 
the kinematics and can now focus on the simpler objects $A_n$.

These color-ordered amplitudes turn out to have an immensely rich structure. 
In particular, they
satisfy several different kind of relations, 
reducing the number of independent $A_n$'s considerably.
The most simple are cyclicity and reversion:
\begin{align}
A_n(1,2,\ldots,n) = A_n(2,3,\ldots,n,1), \qquad A_n(1,2,\ldots,n) = 
(-1)^n A_n(n,n-1,\ldots,1),
\end{align}
The so-called photon-decoupling identity reads
\begin{align}
\sum_{\sigma\in cyclic} A_n(1,\sigma(2),\sigma(3),\ldots,\sigma(n)) = 0,
\end{align}
with the sum running over all cyclic permutations of leg $2,3,\ldots,n$. 
For example, in
the four-point case it reads
\begin{align}
A_4(1,2,3,4) + A_4(1,3,4,2) + A_4(1,4,2,3) = 0.
\end{align}
The photon-decoupling identity reflects the simple fact that we can consider
a Yang-Mills amplitude and replace one of the non-Abelian generators by the
identity matrix. Since it commutes right through all the other generators,
and since the amplitude must vanish, the identity follows.
 
More generally, we have the Kleiss-Kuijf (KK) relations \cite{KK}
\begin{align}
A_n(1,\{\alpha\},n,\{\beta\}) =
(-1)^{n_{\beta}} \!\!\!\!\sum_{\sigma\in \mathrm{OP}(\{\alpha\},
\{\beta^T\})}\!\!\!\! A_n(1,\{\sigma\},n),
\end{align}
where the sum is over ``ordered permutations'', \textit{i.e.} all permutations 
of $\{\alpha\}\cup \{\beta^T\}$
that keep the order of the individual elements belonging to each set fixed. 
Here $n_{\beta}$ is the number of 
elements in $\{\beta\}$, and $\{\beta^T\}$ is the $\{\beta\}$ set 
with the ordering reversed. For example,
\begin{align}
A_6(1,2,3,6,4,5) ={}& A_6(1,2,3,5,4,6) + A_6(1,2,5,3,4,6) + 
A_6(1,2,5,4,3,6) \nonumber \\
& + A_6(1,5,2,4,3,6) + A_6(1,5,4,2,3,6) + A_6(1,5,2,3,4,6),
\end{align}
where we have chosen $\{\alpha\} = \{ 2,3\}$ and $\{\beta\} = \{4,5\}$ 
(and therefore $n_{\beta} = 2$).
Altogether these relations reduce the number of independent 
subamplitudes to $(n-2)!$.

More recently additional relations were discovered: 
the Bern-Carrasco-Johansson (BCJ) relations \cite{BCJ}. 
There are several ways of presenting them, \textit{e.g.}
\begin{align}
0 ={}& s_{12}A_n(1,2,3,\ldots,n) + (s_{12}+s_{23})A_n(1,3,2,4,\ldots,n) + 
(s_{12}+s_{23}+s_{24})A_n(1,3,4,2,5,\ldots,n) \nonumber \\
& + \cdots + (s_{12}+s_{23}+s_{24}+\cdots + s_{2(n-1)})
A_n(1,3,4,\ldots,n-1,2,n),
\label{BCJ}
\end{align}
along with all the relations obtained by permutation of $1,2,\ldots,n$ in 
the above equation. Exploiting these additional relations, 
the number of independent subamplitudes reduces to $(n-3)!$. What is new
here is the fact that the identities involve the external momenta.
Nevertheless, they are universal and hold for all choices of helicity.

The string theory generalization of these relations has also been found 
\cite{BDV,Stieberger}. They follow from monodromy relations in the 
integral representation of string and they reduce to the BCJ-relations in 
the field theory limit. This was in fact the first proof of the
BCJ-relations in the field theory limit. The fact that the basis of
amplitudes is reduced to $(n-3)!$ from $(n-2)!$ follows beautifully
from string theory as a consequence of the need to fix precise three
of the $n$ positions of the external legs in the string theory
amplitude. The origin of this lies in the fact that the Moebius group
is a three-parameter group. It is quite remarkable that this
concept, which would appear to be totally unrelated to QCD, can have
have such an important consequence for QCD amplitudes. Understanding
this in terms of a different representation of the Feynman diagrams
involved is one outstanding question which remains to be answered.
It is possible that the world-line formalism \cite{Strassler} holds
the clue, but the details remain to be worked out. This being said,
it is nevertheless now known how to prove the BCJ-relations using
field theory alone \cite{Feng,Chen:2011jx}. In fact, that proof
does not use any Lagrangian-specific representation, but relies instead
on very general analyticity properties of the $S$-matrix.

\section{Quadratic amplitude relations\label{quad_rel}}

We now come to the main topic of this talk: a new set of identities
among gauge-theory 
amplitudes that have recently been discovered (and proven) in refs.
\cite{BjerrumBohr:2010zb,KLTshort}.
These new identities have a rather different structure than any of the
previously mentioned relations. First of all, they are {\em quadratic} in 
the amplitudes, and second, instead of being helicity-independent,
like all the relations reviewed in the last section, they relate 
amplitudes from different helicity sectors. The fact that the
relations are not linear in amplitudes is particularly surprising; 
almost all of our intuition about gauge theory amplitudes come
from the linear level. This is where Ward Identities live, and this
is where symmetries act in a simple ways. Most unusual,
these quadratic identites can be seen as being dual
to relations that link QCD-amplitudes to {\em gravity amplitudes}. The
fact that they are quadratic is therefore directly related to
the fact that the graviton, being of spin-2, has polarization
tensors that can be factorized into to two spin-1 (gluon) polarization
vectors. Say, for positive helicity: 
$\epsilon^{(+)}_{\mu\nu} \sim \epsilon^{(+)}_{\mu}\epsilon^{(+)}_{\nu}$. 
The appearance of the outer product between the two gluon polarization
vectors is what leads to the quadratic relations. Yet, QCD-amplitude
identities should not involve gravity. How can this be reconciled?  
It turns out that there are ``forbidden'' gravity amplitudes
that vanish. This happens when we consider mixed combinations
$\epsilon^{(+)}_{\mu}\epsilon^{(-)}_{\nu}$ -- what should have been
the scalar component of a gravity amplitude, which vanishes.

Figuratively speaking, then, these new QCD-identities arise as
follows: When we consider appropriate products of like-helicity
amplitudes, we produce a gravity amplitude. When we flip the
helicities of some of the gluons
(which legs we flip will be described in detail below), we instead
get zero. This gives the new identities.

Before presenting the quadratic relations in all their gory detail, 
let us first introduce a useful quantity $\mathcal{S}$, that depends on 
$k$ massless momenta in the following way:
\begin{equation}
{\cal S}[i_1,\ldots,i_k|j_1,\ldots,j_k]_{p_1} \equiv
\prod_{t=1}^k \big(s_{i_t 1}+\sum_{q>t}^k \theta(i_t,i_q)
s_{i_t i_q}\big) \label{Sdef}\,,
\end{equation}
where $\{i_1,\ldots,i_k\}$ and $\{j_1,\ldots,j_k\}$ are two arbitrary 
orderings of $k$ momenta. We have defined as usual 
$s_{ij} \equiv (p_i + p_j)^2$. Furthermore,
$\theta(i_a,i_b)$ is zero if $i_a$ comes  sequentially before 
$i_b$ in $\{j_1,\ldots,j_k\}$ and unity if it comes after.

The definition of $\mathcal{S}$ may sound complicated, in particular
dues to the curious definition of the ``step function'' $\theta(i_a,i_b)$.
It is therefore useful to write down a few
explicit examples of $\mathcal{S}$ that will help in understanding
how it is constructed:
\begin{align}
\mathcal{S}[2|2]_{p_1} = s_{12}, \qquad \mathcal{S}[23|23]_{p_1} = 
s_{12}s_{13}, \qquad
\mathcal{S}[234|324]_{p_1} = (s_{12}+s_{23})s_{13}s_{14}.
\end{align}

This $\mathcal{S}$ function turns out to have quite a few very nice 
properties. We will not go through all of them here, they can be found
in $\cite{BjerrumBohr:2010hn}$ (along with its string theory generalization,
which turns out to preserve these nice properties),
but we wish to stress its close connection to the BCJ-relations. 
This is most manifest through
\begin{align}
0 = \sum_{\alpha\in S_{n-2}} \mathcal{S}[\beta(2,n-1)|
\alpha(2,n-1)] A_n(1,\alpha(2,n-1),n),
\label{SA}
\end{align}
where we have introduced the simplifying notation $\alpha(2,n-1) \equiv 
\{\alpha(2),\alpha(3),\ldots,\alpha(n-1)\}$ for
permutations of leg $2,3,\ldots,n-1$, and $\beta(2,n-2)$ is just some 
arbitrary permutation of the same legs. The sum is
over all $\alpha$ permutations. Eq.~\eqref{SA} is indeed precisely 
satisfied due to the BCJ-relations in eq.~\eqref{BCJ}! Conversely,
the system of equations one obtains from choosing different $\beta$ 
permutations in eq.~\eqref{SA} can be used to reduce the basis
to $(n-3)!$ \cite{BjerrumBohr:2010hn}. In this way we see that
eq.~\eqref{SA} is a neat and compact reformulation of the BCJ-relations 
in terms of our $\mathcal{S}$ function, the momentum kernel.

Since the quadratic relations which we are soon going to present
are helicity-dependent, we will also introduce the following
short-hand notation. We denote $A_n^k$ an N$^k$MHV amplitude, 
\textit{i.e.} $A_n^k$ is an $n$-point subamplitude with $2+k$
negative helicity gluons. Of course, to have non-vanishing amplitudes 
$k\in\{0,1,\ldots,n-4\}$.
We do not care about the exact helicity configuration, just which 
helicity \textit{sector} it belongs to.

With this we can now write down the quadratic relations. Assuming 
$k\neq h$ the following identity is satisfied
\begin{equation}
0  =  \sum_{\gamma,\beta\in S_{n-3}}
A_n^k(n-1,n,\gamma(2,n-2),1) \mathcal{S}[\gamma(2,n-2)|\beta(2,n-2)]_{p_1}
A_n^h(1,\beta(2,n-2), n-1,n),
\label{prod_vanish}
\end{equation}
where the sum is over all permutations of $\gamma$ and $\beta$.

Note that in each of the $A_n^k$ ($A_n^h$) amplitudes in the sum 
it is the same $k+2$ ($h+2$) legs that have flipped helicity.
These are relations among subamplitudes living in different helicity sectors. 
They are \textit{not} just satisfied due to the BCJ-relations, 
which one might have thought from the seemingly close analogy to
eq.~\eqref{SA}. The fact that the identities have quite different
content is clearly pointed out from our requirement 
$k\neq h$ on the helicity structure, whereas
the BCJ-relations are helicity-independent. 
We can schematically express eq.~\eqref{prod_vanish} as

\begin{eqnarray}
\hspace{1cm}
\mbox{\fontsize{20}{24}\selectfont $0=\sum$}\hspace{1cm}
\parbox{50mm}{
\begin{fmffile}{ASA1}
     \begin{fmfgraph*}(80,80)
     \fmfsurroundn{v}{8}
\fmf{plain}{v1,c1}
\fmf{plain}{v2,n1}
\fmf{plain}{n1,c1}
\fmf{phantom}{v3,c1}
\fmf{plain}{v4,m1}
\fmf{plain}{m1,c1}
\fmf{plain,label=$\hspace{-1.5cm}n-1$}{v5,c1}
\fmf{plain}{v6,m2}
\fmf{plain}{m2,c1}
\fmf{phantom}{v7,c1}
\fmf{plain}{v8,n2}
\fmf{plain}{n2,c1}

\fmfv{decor.shape=circle,decor.filled=30,decor.size=0.30w,label=$A_n^k$,label.dist=-6}{c1}
\fmffreeze
     \fmf{dots,left=0.5,tension=0.2}{n1,n2}

\fmflabel{$n$}{v4}
\fmflabel{1}{v6}

\end{fmfgraph*}
\end{fmffile}
}
\hspace{-2cm}
\mbox{\fontsize{16}{19}\selectfont $\times$}
\quad
\mbox{\fontsize{20}{24}\selectfont $\mathcal{S}$}
\quad
\mbox{\fontsize{16}{19}\selectfont $\times$}\:\:
\parbox{50mm}{
\begin{fmffile}{ASA2}
     \begin{fmfgraph*}(80,80)
     \fmfsurroundn{v}{8}
\fmf{plain}{v1,c1}
\fmf{plain}{v2,n1}
\fmf{plain}{n1,c1}
\fmf{phantom}{v3,c1}
\fmf{plain}{v4,m1}
\fmf{plain}{m1,c1}
\fmf{plain}{v5,c1}
\fmf{plain}{v6,m2}
\fmf{plain}{m2,c1}
\fmf{phantom}{v7,c1}
\fmf{plain}{v8,n2}
\fmf{plain}{n2,c1}

\fmfv{decor.shape=circle,decor.filled=30,decor.size=0.30w,label=$A_n^h$,label.dist=-6}{c1}
\fmffreeze
     \fmf{dots,right=0.5,tension=0.2}{m1,m2}

\fmflabel{$n-1$}{v2}
\fmflabel{1}{v1}
\fmflabel{$n$}{v8}

\end{fmfgraph*}
\end{fmffile}
}
\end{eqnarray}

We re-emphasize that when helicities are not flipped, the same right
hand side of this equation produces instead a graviton amplitude!
The explicit link between the two sets of relations has been
further elaborated on in ref. \cite{Tye}.

Although eq.~\eqref{prod_vanish} is probably one of the nicest forms 
of the general $n$-point case,
it contains a quite huge number of terms, namely $[(n-3)!]^2$. 
Using BCJ-relations one can write eq.~\eqref{prod_vanish}
in other, but completely \textit{equivalent}, 
ways \cite{BjerrumBohr:2010yc}. In general we can write
the relations as
\begin{align}
0  = &\!\!\! \sum_{\sigma\in S_{n-3}}\sum_{\alpha\in
S_{j-2}}\sum_{\beta\in S_{n-1-j}}
\!\!\! A_n^k(\alpha(\sigma(2,j-1)),1,n-1,\beta(\sigma(j,n-2)),n)
\mathcal{S}[ \alpha(\sigma(2,j-1))|\sigma(2,j-1)]_{p_1}\nonumber \\
&  \hspace{2cm}\times \mathcal{S}[\sigma(j,n-2)
|\beta(\sigma(j,n-2))]_{p_{n-1}}A_n^h(1,\sigma(2,j-1),\sigma(j,n-2), n-1,n)
\label{gen_vanish}\,,
\end{align}
for \textit{any} $j = 2,3,\ldots,n-1$. Eq.~\eqref{prod_vanish} is therefore 
just a special case of eq.~\eqref{gen_vanish}, namely the one with $j=n-1$.
However, we could just as well choose $j=[n/2]+1$ in which case the number 
of terms in eq.~\eqref{gen_vanish} reduces to $(n-3)!([n/2]-1)!([n/2]-2)!$

Finally, we provide yet another way of writing these relations. 
This form has a higher degree of
manifest crossing symmetry between the different external legs. 
However, it requires the introduction of a regularization. To this end,
let us assume we make the
following shift in the momentum $p_1$ of leg 1 and momentum $p_n$ of
leg $n$:
\begin{align}
p_1 \rightarrow p_1' \equiv p_1 -xq, \qquad p_n \rightarrow p_n' 
\equiv p_n + xq,
\end{align}
where $x$ is some arbitrary parameter and $q$ a four-vector satisfying 
$q^2 = p_1\cdot q = 0 \neq p_n\cdot q$.
This preserves overall energy-momentum conservation, 
keeps $p_1'$ on-shell, but makes ${p_n'}^2 = s_{1'2\ldots n-1}\neq 0$.
After this regularization the quadractic relations can be written as
\begin{align}
0 = \lim_{x\rightarrow 0} \sum_{\gamma,\beta} 
\frac{A_n^k(n',\gamma(2,n-1),1') 
\mathcal{S}[\gamma(2,n-1)|\beta(2,n-1)]_{p_1'}
A_n^h(1',\beta(2,n-1),n')}{s_{1'2\ldots n-1}}.
\label{sing_vanish}
\end{align}
We see that as $x\rightarrow 0$ the denominator goes to zero, 
but due to eq.~\eqref{SA} the numerator is {\em also} going to zero.
Indeed, both the numerator and denominator vanish at the same, 
and eq.~\eqref{sing_vanish} is just another way
of representing the quadratic relations \cite{KLTshort}.

Although this form may not be as practically convenient as 
eq.~\eqref{gen_vanish} it turns out to be important for the
proof of the quadratic relations, even in the form of eq.~\eqref{gen_vanish}.

\subsection{Examples}

To get a better feel for eq.~\eqref{gen_vanish} 
(and eq.~\eqref{prod_vanish}), let us write out
some explicit examples. For four points the relations are trivially 
satisfied. Eq.~\eqref{prod_vanish} takes the form
\begin{align}
0 = s_{12}A_4^k(3,4,2,1)A_4^h(1,2,3,4).
\end{align}
But if $k\neq h$ at least one of the amplitudes must have three or four of 
the same helicity and therefore vanish
all by its own due to the MHV-rule. (Alternatively, here is a quite
unusual proof of that MHV-rule).
However, already at five points we start getting non-trivial cancellations
among different amplitudes that do not vanish individually. 
In this case eq.~\eqref{prod_vanish} is
\begin{align}
 0={}&
s_{12}A_5^h(1,2,3,4,5)\big[ s_{13}A_5^k(4,5,2,3,1)
+ (s_{13}+s_{23})A_5^k(4,5,3,2,1)\big] \nonumber \\
& +s_{13}A_5^h(1,3,2,4,5)\big[s_{12}A_5^k(4,5,3,2,1) + (s_{12}
+s_{23})A_5^k(4,5,2,3,1)\big],
\end{align}
or using eq.~\eqref{gen_vanish} with $j=3$
\begin{align}
 0= s_{12}s_{34}A_5^h(1,2,3,4,5)A_5^k(4,3,5,2,1)  
+s_{13}s_{24}A_5^h(1,3,2,4,5) A_5^k(4,2,5,3,1).
\end{align}
Taking $(h,k)=(0,1)$ (or $(1,0)$) we have relations with non-vanishing 
subamplitudes. For instance,
\begin{align}
 0={}& s_{12}s_{34}A_5(1^-,2^-,3^+,4^+,5^+)
A_5(4^+,3^-,5^+,2^-,1^-) \nonumber \\
& +s_{13}s_{24}A_5(1^-,3^+,2^-,4^+,5^+) A_5(4^+,2^-,5^+,3^-,1^-).
\end{align}
Finally, let us give the explicit expression for six points 
(using eq.~\eqref{gen_vanish} with $j=4$)
\begin{align}
0 ={}& s_{12}s_{45}A_6^h(1,2,3,4,5,6) \big[   s_{13} A_6^k(5,4,6,2,3,1)
 +(s_{13}+s_{23}) A_6^k(5,4,6,3,2,1) \big]  + \mathcal{P}(2,3,4).
\label{6ptex}
\end{align}
Here we start having relations between ``real'' NMHV and MHV amplitudes 
($(h,k)=(0,1)$).

It is clear from this discussion that 
if we know the amplitudes for some helicity sector, we can plug them into 
the quadratic relations and thereby get linear relations between amplitudes 
in {\em another} helicity sector. In particular, one can choose
the simple MHV amplitudes for one of them and get linear relations between 
the more complicated N$^h$MHV amplitudes. Choosing the $A_n^k$ amplitudes 
to be MHV, and using a standard
spinor-helicity notation, 
one gets the general relation \cite{Feng:2010br,Feng:2010hd}
\begin{align}
0 = \sum_{\beta\in S_{n-3}} \prod_{i=2}^{n-2} 
\lbrack \beta(i)| \beta(i+1) + \beta(i+2) +\cdots +\beta(n-1)|n\rangle
A_n^h(1,\beta(2,n-2),n-1,n),
\end{align}
for $h>0$. As an example, consider the six-point case where $A_6^h$ 
is some NMHV amplitude
\begin{align}
0 = \lbrack 2| 3+4+5|6\rangle \lbrack 3| 4+
5|6\rangle \lbrack 4|5|6\rangle A_6^h(1,2,3,4,5,6) + \mathcal{P}(2,3,4).
\end{align}

One might wonder from these considerations 
what happens when $k=h$. In this case we do not have 
zero on the left-hand-side of eq.~\eqref{gen_vanish} (or 
eq.~\eqref{sing_vanish}),
but instead this is precisely where we get a tree-level gravity amplitude!
That is, when the helicities in the two amplitudes are the same
one has the famous Kawai-Lewellen-Tye (KLT) relations \cite{KLT}. 
Surprisingly, the vanishing of these quadratic relations was
needed as an essential input for the field theoretical proof of the 
KLT-relations \cite{KLTshort,BjerrumBohr:2010yc}. One can unify
these relations by consider the KLT-relations between $\mathcal{N}=4$ 
supersymmetric Yang-Mills and $\mathcal{N}=8$ supergravity. 
In this picture the quadratic vanishing relations appear as a consequence of
violation of $R$-symmetry \cite{Feng:2010br,Tye}.

\section{The proof\label{proof}}

The reader may be more interest in the content of these QCD-identities
than in their proof. Nevertheless, let us here just sketch the proof
of these new quadratic relations. All we need to know are some
analytical properties of subamplitudes. For full details of the proof we
refer to \cite{BjerrumBohr:2010yc}.

We choose to prove the identities by induction. We thus assume that we have
verified the quadratic identities up to $n-1$
points (they are easy to verify in the low-point cases) and now want to 
show this implies the relation at $n$ points,
\textit{i.e.} show that
\begin{align}
X_n  =  \sum_{\gamma,\beta\in S_{n-3}}
A_n^k(n-1,n,\gamma(2,n-2),1) \mathcal{S}[\gamma(2,n-2)|\beta(2,n-2)]_{p_1}
A_n^h(1,\beta(2,n-2), n-1,n)
\end{align}
is also zero (when $k\neq h$). Here we restrict ourselves to the form in
eq.~\eqref{prod_vanish}, since it is
completely equivalent to all the forms contained in 
eq.~\eqref{gen_vanish}, see \cite{BjerrumBohr:2010yc}. Showing the
new identities in this particular form is therefore equivalent
to showing them all.

We now make a so-called BCFW-shift in the legs $1$ and
$n$, and consider the following contour integral \cite{BCFW}
under the assumption that the boundary contribution is zero,
\begin{align} 0 =
\oint \frac{dz}{z}X_n(z) = X_n(0) +
\sum (\mathrm{residues\:\:for}\:\:z\neq 0)\,.
\end{align}
Since the residues are calculated from the poles of $X_n$, and the poles
only appear in the subamplitudes, there are two different kind of terms 
to consider:
\begin{itemize}
\item[(A)] The pole appears in only one of the
    amplitudes $A_n^k$ or $A_n^h$.
\item[(B)] The pole is present in both
    amplitudes $A_n^k$ and $A_n^h$.
\end{itemize}

Using the properties of the $\mathcal{S}$ function one find that the terms in
category (A) always vanish due to BCJ-relations (in the form of 
eq.~\eqref{SA}).

The terms in category (B) are a bit more tricky. However, these will 
factorize into products of lower-point quadratic relations 
where at least one of them will satisfy the requirement that
the amplitudes belong to different helicity sectors.
For this part it is important that we know that the quadratic relations
can be written in the form of eq.~\eqref{sing_vanish} since it will be 
products of quadratic relations with
one in the form of eq.~\eqref{prod_vanish} and one in the form of 
eq.~\eqref{sing_vanish}. For this step we
would of course first need to show that eq.~\eqref{sing_vanish} is 
always satisfied. This can be done from the  beginning by following
the exact same procedure. Actually, in that case the residues will 
always factorize into quadratic
relations in the form of eq.~\eqref{sing_vanish} and thereby make 
the induction proof direct.

With this procedure we find that \textit{all} residues vanish and 
therefore also $X_n \equiv X_n(0) = 0$.

\section{Conclusion\label{con}}
We have discussed in some detail the quadratic identites among 
Yang-Mills amplitudes that have recently been found \cite{BjerrumBohr:2010zb}.
Along with several explicit examples we have also sketched our proof of 
these new identities.  
What is surprising about these QCD-relation is that it seems one could 
not possibly have arrived at them without knowing that there is also a 
way to construct gravity amplitudes out of quadratic combinations of
not-flipped color-orderes amplitudes. We have nevertheless succeeded
in proving them directly using only self-contained quantum field theory 
tools, without any reference to gravity amplitudes.
Further progress in this direction seems possible. 
It would clearly be interesting to 
investigate more identities 
at one and possibly multi-loop level. 
This seems a promising avenue for future studies.
Some steps have already been taken in this direction \cite{BjerrumBohr:2010zb}.

\end{document}